\documentstyle[12pt]{article}  

\begin {document}
\twocolumn[
{\centering
\begin{large} ~
\end{large} \\[0.5cm]
~ \\[0.5cm]
Mechanism of Magnetism in Stacked Nanographite\\[0.5cm]
ETL$^1$, Tokyo Institute of Technology$^2$:
K. Harigaya$^{1,2}$, N. Kawatsu$^2$, T. Enoki$^2$
\\[0.5cm] }]

Nanographite systems, where graphene\\ sheets of the orders of 
the nanometer size are stacked, show novel magnetic properties, 
such as, spin-glass like behaviors [1], and the change of 
ESR line widths while gas adsorptions [2]. Recently, it 
has been found [3] that magnetic moments decrease with the 
decrease of the interlayer distance while water molecules 
are attached physically.

In this paper, we consider the mechanisms of antiferromagnetism 
using the Hubbard-type model with the interlayer hopping
$t_1$ and the onsite repulsion $U$ [4].

First, the active functional groups are simulated with introducing
a site potential $E_s$ [5] at edge sites.  When $E_s > 0$,
the site potential means the electron attractive groups.
When $E_s < 0$, the electron donative groups are simulated
because of the increase of the electron number at the
site potentials.  Here, we take $E_s = -2t$, and one
additional electron per layer is taken account.
We find that the total magnetization is a decreasing function
of $t_1$.  The decrease is faster when the
edge atom with the site potential is neighboring to the
site with the interaction $t_1$, and thus the localized
character of the magnetic moment can be affected easily.
The decease of magnetization by the
magnitude $30-40$\% with the water molecule attachment [3]
may correspond to this case.

Next, we look at the magnetism of stacked ``triangulenes".
The ``triangulene" has nine hexagonal rings [6].  The Lieb's theorem 
[7] says that the total spin $S_{\rm tot}$ of the repulsive 
Hubbard model of the A-B bipartite lattice is 
$S_{\rm tot}=\frac{1}{2}| N_A - N_B |$, where $N_A$ and $N_B$ 
are the numbers of A and B sites.  We find $S_{\rm tot} = 1$
for the single triangulene.  The total magnetic moment per layer 
for the A-B stacking with the vertical shift is a decreasing function with 
respect to $t_1$.  As we discuss in detail [4], there appear 
strong local magnetic moments at the zigzag edge sites, 
and they give rise dominant contributions to the magnetism 
of each layer.  In the triangulene case, most of the edge
sites are neighboring to the sites with the interaction $t_1$.  
The interactions of the edge sites with the neighboring layers 
are strong, and the itinerant characters 
of electrons become larger as increasing $t_1$.  Therefore, 
the magnetic moment is a decreasing function.

The present two calculations agree with the experiments, 
qualitatively.  We can explain the decrease of magnetism
in the process of adsorption of molecules [3].
Thus, the open shell electronic structures due to 
the active side groups and/or the geometrical origin 
are candidates which could explain the exotic magnetisms.

\mbox{}

\noindent
$[1]$ Y. Shibayama {\sl et al.}, 
Phys. Rev. Lett. {\bf 84}, 1744 (2000).\\
$[2]$ N. Kobayashi {\sl et al.},
J. Chem. Phys. {\bf 109}, 1983 (1998).\\
$[3]$ N. Kawatsu {\sl et al.},
Meeting Abstracts of the Physical Society of Japan
{\bf 55} Issue 1, 717 (2000).\\
$[4]$ K. Harigaya, J. Phys.: Condens. Matter {\bf 13}, 
1295 (2001); cond-mat/0010043; cond-mat/0012349.\\
$[5]$ K. Harigaya, A. Terai, Y. Wada, and K. Fesser,
Phys. Rev. B {\bf 43}, 4141 (1991).\\
$[6]$ G. Allinson, R. J. Bushby, and J. L. Paillaud,
J. Am. Chem. Soc. {\bf 115}, 2062 (1993).\\
$[7]$ E. H. Lieb, Phys. Rev. Lett. {\bf 62},
1201 (1989); {\sl ibid.} {\bf 62}, 1927 (1989).\\

\end{document}